\documentclass[twocolumn,aps,pre,preprintnumbers,amsmath,amssymb,nofootinbib,floatfix]{revtex4-1}

\usepackage{graphicx,color}
\usepackage{caption}
\usepackage{subcaption}
\usepackage{amsmath}
\usepackage{epstopdf}
\usepackage{soul}
\newcommand{\D}{\mathrm{d}}
\newcommand{\e}{\mathrm{e}}
\newcommand{\half}{\frac{1}{2}}
\newcommand{\be}{\begin{equation}}
\newcommand{\ee}{\end{equation}}
\newcommand{\bea}{\begin{eqnarray}}
\newcommand{\eea}{\end{eqnarray}}
\newcommand{\ba} {\begin{align} }
\newcommand{\ea} {\end{align} }
\newcommand{\eps}{\varepsilon}

\newcommand{\kbt}{k_{\mathrm{B}}T}

\newcommand{\lb}{l_{{\rm B}}}
\newcommand{\ld}{\lambda_{{\rm D}}}

\newcommand{\vecr}{{\bf r}}

\newcommand {\eeff} {\varepsilon_{{\rm eff}}}
\newcommand {\keff} {\kappa_{{\rm eff}}}
\newcommand {\leff} {\lambda_{{\rm eff}}}

\newcommand{\ra}[1]{\textcolor{black}{#1} } 

\begin{document}


\title{Bjerrum Pairs in Ionic Solutions: a Poisson-Boltzmann Approach}

\author{Ram M. Adar$^{1}$, Tomer Markovich$^{1,2}$, David Andelman$^1$}
\email{andelman@post.tau.ac.il}
\affiliation{$^1$Raymond and Beverly Sackler School of Physics and Astronomy\\ Tel Aviv
University, Ramat Aviv, Tel Aviv 69978, Israel\\
$^2$DAMTP, Centre for Mathematical Sciences, University of Cambridge\\ Cambridge CB3 0WA, United Kingdom}


\begin{abstract}
Ionic solutions are often regarded as fully dissociated ions dispersed in a polar solvent. While this picture holds for dilute solutions, at higher ionic concentrations, oppositely charged ions can associate into dimers, referred to as {\it Bjerrum pairs}. We consider the formation of such pairs within the nonlinear Poisson-Boltzmann framework, and investigate their effects on bulk and interfacial properties of electrolytes. Our findings show that pairs can reduce the magnitude of the dielectric decrement of ionic solutions as the ionic concentration increases. We describe the effect of pairs on the Debye screening length, and relate our results to recent surface-force experiments. Furthermore, we show that Bjerrum pairs reduce the ionic concentration in bulk electrolyte and at the proximity of charged surfaces, while they enhance the attraction between oppositely charged surfaces.
\end{abstract}

\maketitle

\section{Introduction}
\label{sec1}
Ionic solutions are ubiquitous in electrochemical, colloidal and biological systems. The solution properties are determined by the interplay between the ion mixing entropy and their electrostatic interaction~\cite{Israelachvily,VO,David95}. An important length emerging from this interplay is the Bjerrum length, $\lb=e^{2}/\left(4\pi\eps\kbt\right)$,  where $e$ is the  electronic unit charge, $\eps$ the dielectric constant of the solution, and $\kbt$  the thermal energy. At this length, the Coulombic interaction between two unit charges is equal to the thermal energy. For water, $\eps\approx78\,\eps_{0}$, where $\eps_{0}$ is the vacuum permittivity, and the Bjerrum length is equal to about 0.7\,nm, at room temperature.

When the Bjerrum length is comparable with the lattice spacing, it is favorable for a salt crystal ({\it e.g.,} NaCl) to dissociate, forming an ionic aqueous solution. In such solutions, another important length scale naturally emerges. This is the Debye screening length, $\ld$, that was introduced in the 1920's~\cite{DH} by Debye and H\"uckel (DH). For fully dissociated monovalent salt with bulk concentration $n_b$, the Debye screening length is given by
\begin{equation}
\ld=\frac{1}{\sqrt{8\pi\lb n_{b}}}\, ,
\label{ld}
\end{equation}
and at distances larger than $\ld$, the electric field induced by an ion is exponentially screened.

The apparent dichotomy between salt crystal and fully dissociated free ions in solution is an over-simplification, especially for concentrated solutions, where complexation of ions has been proposed as an alternative structure already in the early works of Bjerrum~\cite{Bjerrum26}. Bjerrum  postulated that oppositely charged ions in solution can associate into ionic pairs (dipoles), which are nowadays referred to as {\it Bjerrum pairs}. These pairs were shown, for example,  to play an important role on the critical behavior of ionic solutions~\cite{Fisher93,Levin96}. In addition, they arise naturally in two-component hardcore plasma~\cite{Netz00}.

The formation of ionic pairs has a two-fold effect on the screening length, $\ld\sim (\eps/n_b)^{1/2}$. First, pairs reduce the concentration of free ions that participate in the screening to a value lower than $n_b$~\cite{Zwanikken09}. Second, pairs increase the solvent permittivity, $\varepsilon$, due to their permanent dipolar moment. \ra{Furthermore, free ions are known to lower the dielectric constant of ionic solutions~\cite{Hasted48,Barthel95}. This phenomenon was recently attributed to correlation effects~\cite{Levy12,Levy13} and may also be related to the exclusion of water dipoles, as is described below. In this picture, ionic pairs can increase the dielectric constant by excluding less water dipoles.}

The simple remarks above are related to recent surface-force experiments~\cite{Smith16} conducted on ionic liquids and solutions at relatively high ionic concentrations, (up to about 4\,M). The screening length fitted from the force profiles between two surfaces was shown to be non-monotonic as function of the ionic concentration.  It first decreases for low ionic concentrations, and then anomalously increases for higher values.

In this paper, we systematically incorporate the formation of Bjerrum pairs into the Poisson-Boltzmann (PB) theory. We demonstrate that pairs have a qualitative effect on bulk electrolyte properties. Our model predicts that a non-negligible fraction of salt ions associate into pairs. \ra{Considering both free ions and pairs,} we are able to account for the nonlinear behavior of the dielectric decrement of ionic solutions on the mean field (MF) level. Furthermore, the screening length is found to be qualitatively modified by Bjerrum pairs, depending on the relative strength of the dipolar moment of the ion pairs, as compared to that of the solvent. Finally, we obtain the effect of Bjerrum pairs on the counterion profiles next to charged surfaces, and on the osmotic pressure between two oppositely charged surfaces.

The outline of our paper is as follows. The model is formulated in Section~\ref{sec2}, where we calculate the bulk concentration of Bjerrum pairs. In Section~\ref{sec3}, we describe how ion pairs modify the dielectric constant and screening length. Next, in Section~\ref{sec4}, we present results for the ionic profiles at the proximity of a charged surface and elaborate on the local effects of the pair formation, and in Section~\ref{sec5}, we solve the two-plate problem and calculate the corresponding osmotic pressure. Finally, Section~\ref{sec6} offers a discussion that includes a comparison with previous models and relevant experiments.

\section{Model}
\label{sec2}

Consider an aqueous solution with added electrolyte. For simplicity, we assume a monovalent electrolyte with cations/anions of unit charge $\pm e$. The two ionic species have the same bulk concentration, $n_{b}$, satisfying electro-neutrality. The solvent (water) molecules are modeled as dipoles of concentration $n_{w}$ and permanent dipole moment $p_{w}$. We further assume that a fraction of the cations and anions can associate into Bjerrum pairs, modeled as dipoles with moment $p=be$, and the length $b$ corresponds to a typical separation between the paired ions.  As a result of the association, the total number of ions partitions into free ions of concentration $n_{s}$, and ion pairs (dipoles) of concentration $n_{p}$, satisfying $n_{s}+n_{p}=n_{b}$. The two ionic states (free ions and pairs) are in chemical equilibrium that determines the $n_s$ and $n_p$ values for a given $n_b$, as is described below.

\subsection{Free-ion and pair concentrations}
\label{bulk_concentration}
The concentrations, $n_{s}$ and $n_{p}$, can be calculated, for example, via a lattice-gas model. We model the bulk solution as a cubic lattice with a unit cell of  volume $a^3$. Each cell is occupied by either solvent,  a cation, an anion, or a Bjerrum pair.  The ion-pair association energy is $-J$, which accounts for the electrostatic attraction and, possibly, short-ranged ion-specific interactions. \ra{As the electrostatic attraction governs in such small length scales, we restrict ourselves hereafter to positive $J$ values, $J>0$. Note that the limit of zero ion pairs corresponds to the limit $J\to -\infty$.} On the MF level, no additional interactions are considered in the bulk.

For simplicity, we assign the same lattice size to the solvent, cations, anions, and ion pairs, implying that all species occupy a similar volume. Note that this volume is not necessarily the bare ionic size. Water molecules form hydration shells around ions and dipoles and swell their effective volume~\cite{Israelachvily}. Hence, the length $a$ does not correspond to the ionic diameter but to the diameter of the hydrated ion.

Without any interaction between unit cells, the grand-canonical partition per cell is given by
\be
\label{eq2}
Z=1+\Lambda_{+}+\Lambda_{-}+\Lambda_{+}\Lambda_{-}\e^{\beta J},
\ee
where $\beta=1/(\kbt)$ is the inverse thermal energy and the fugacities, $\Lambda_{\pm}=\exp\left(\beta\mu_{\pm}\right),$ depend on the chemical potentials, $\mu_{\pm}$. For convenience, the chemical potential of the solvent (water) is set to zero. This is possible, as the number of solvent molecules, $n_w$, is determined from the condition,
\be
\label{eq0}
a^3\left(n_{w}+2n_{s}+n_{p}\right)=1.
\ee

As Eq.~(\ref{eq2}) is symmetric under the exchange $\Lambda_{-}\leftrightarrow\Lambda_{+}$, the two fugacities are equal, $\Lambda_{+}=\Lambda_{-}\equiv\Lambda$. They are  related to the bulk concentration by $2\phi=\Lambda
\partial\ln Z/\partial\Lambda$, where $\phi=n_{b}a^{3}$ is the average number of cations/anions in a unit cell, ranging
from $\phi=0$ for pure solvent to $\phi=1$, where all lattice sites are occupied by Bjerrum pairs. It then follows that
\begin{align}
\label{eq3}
\phi=\left(n_s+n_p\right)a^3=\frac{1}{1+2\Lambda+\Lambda^{2}\e^{\beta J}}\left(\Lambda+\Lambda^{2}\e^{\beta J}\right),
\end{align}
where the first term on the right-hand-side is equal to $n_s a^3$ and the second one is $n_p a^3$.

\begin{figure*}[ht]
\centering
\begin{subfigure}[b]{0.45\textwidth}
\includegraphics[width=0.85\textwidth]{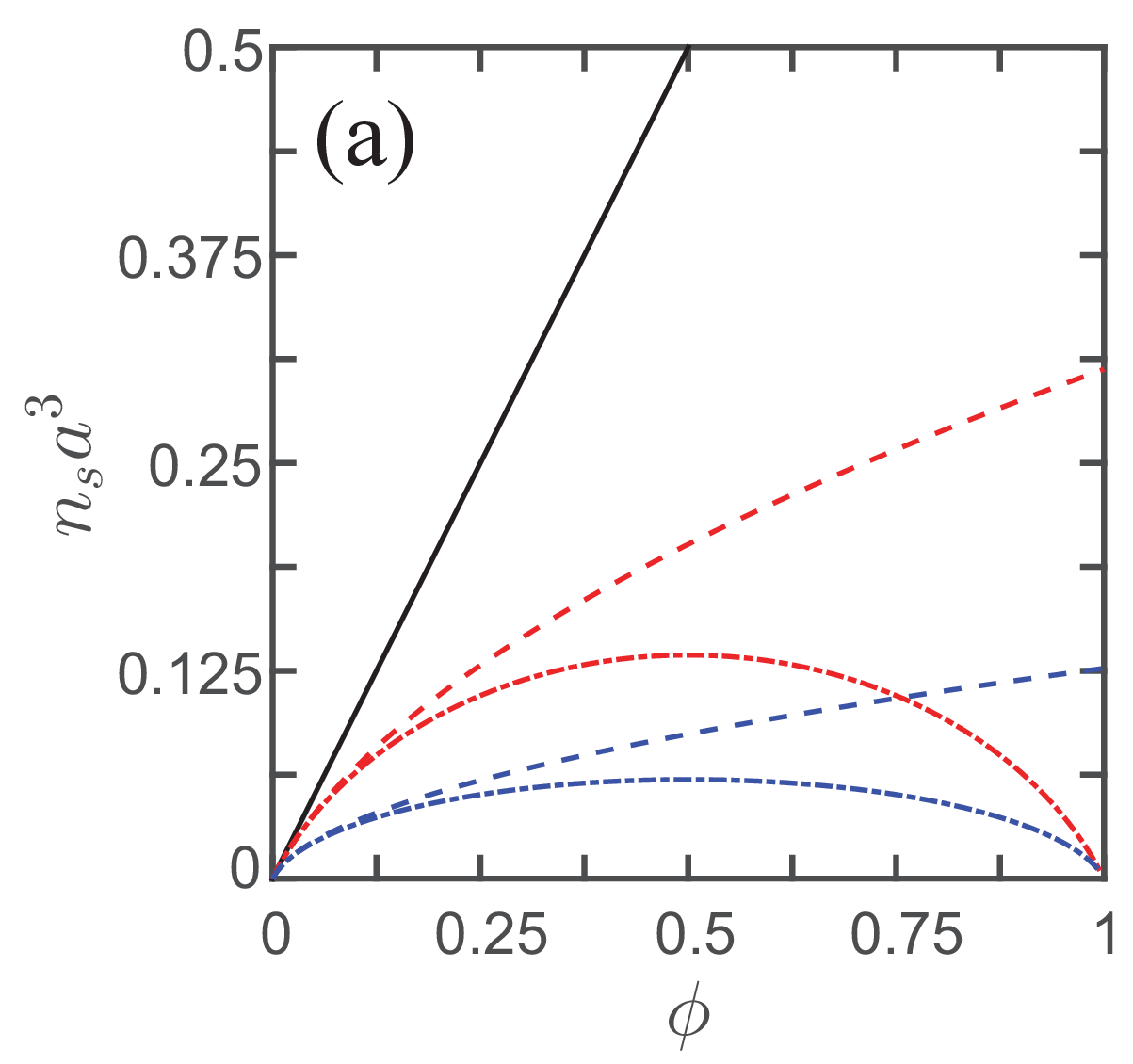}
\end{subfigure}
\begin{subfigure}[b]{0.45\textwidth}
\includegraphics[width=0.85\textwidth]{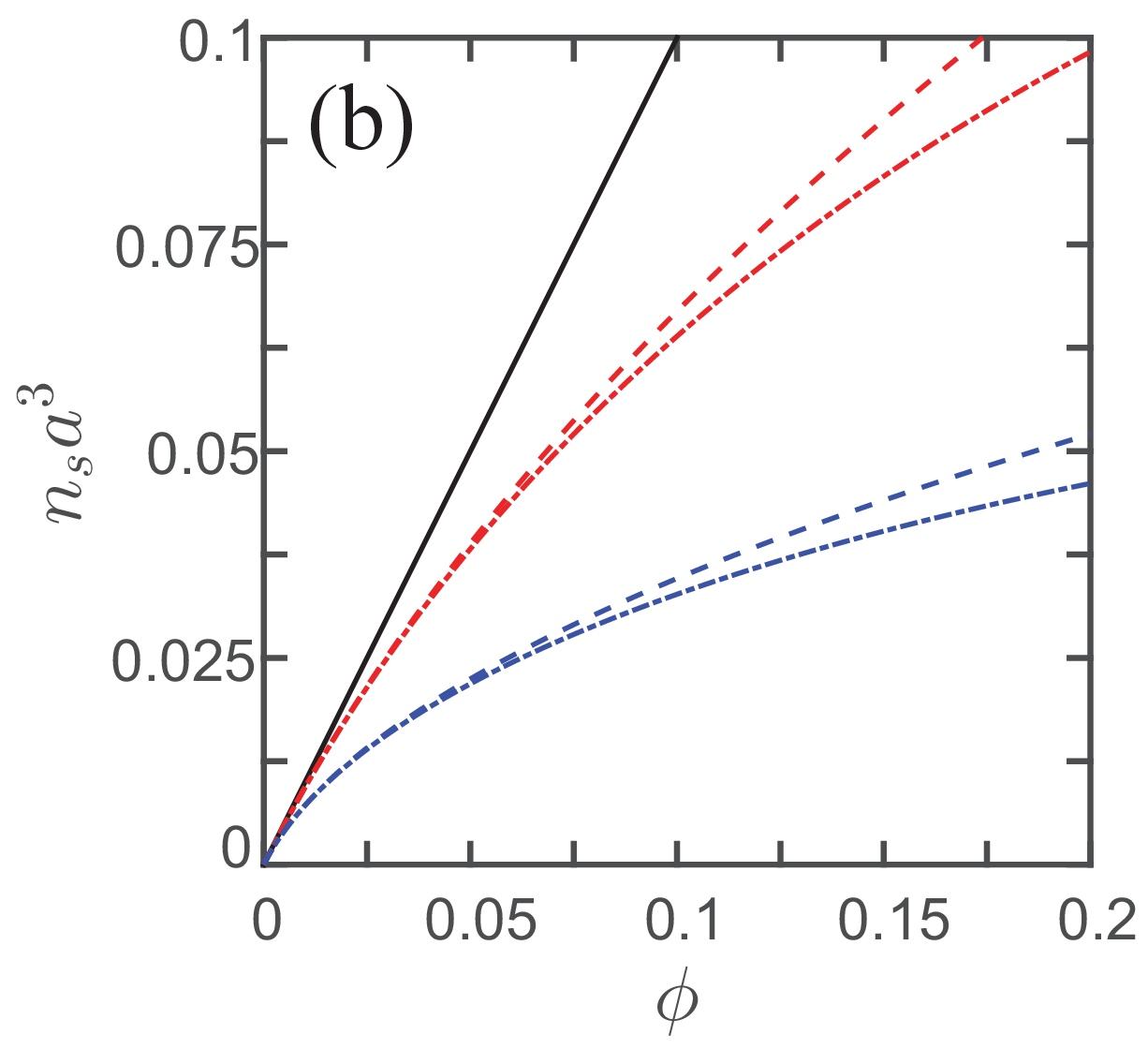}
\end{subfigure}
\caption{Dimensionless free-ion concentration, $n_{s}a^{3}$, as function of $\phi=n_{b}a^{3}$, where $a$ is the lattice size. The fully dissociated concentration without any dipoles ($n_{s}{=}n_{b}$) is plotted as a black solid line. The exact result of Eq.~(\ref{eq7}) is plotted as dot-dashed lines and the approximated  one ($\Lambda\ll1$) of Eq.~(\ref{eq4}) as dashed lines, for $J=2\kbt$ (red) and $J=4\kbt$ (blue). Part (a) shows the dependence for the entire $0\le \phi\le 1$ range, while (b) is a blow-up of the $0\le \phi\le 0.2$ range.}
\label{fig1}
\end{figure*}

Solving Eqs.~(\ref{eq3}) leads to the following expression for $\Lambda(\phi)$:
\be
\label{eq7}
\Lambda=\frac{-\left(1-2\phi\right)+\sqrt{\left(1-2\phi\right)^{2}+4\e^{\beta J}\phi\left(1-\phi\right)}}{2\e^{\beta J}\left(1-\phi\right)}.
\ee
From Eqs.~(\ref{eq3}) and (\ref{eq7}), we obtain the dependence of the free-ion bulk concentration on $\phi$, $n_{s}=n_{s}(\phi)$, as is plotted in Fig.~\ref{fig1}.

The limit of $\Lambda\ll 1$ corresponds to dilute ionic solutions ($\phi\ll 1$), where steric effects are negligible and Eq.~(\ref{eq7}) simplifies to
\be
\label{eq4}
n_{s}a^{3}\simeq\Lambda\simeq\frac{-1+\sqrt{1+4\phi\e^{\beta J}}}{2\e^{\beta J}}.
\ee
The dipolar concentration is then given by $n_{p}=a^{3}n_{s}^2\exp(\beta J)$. The result of Eq.~(\ref{eq4}) can also be obtained straight-forwardly~\cite{Zwanikken09} by introducing a kinetic constant for the reaction, $K=a^{-3}\exp(-\beta J)$, satisfying $K=\left[n_{s}\right]^{2}/\left[n_{p}\right]$. However, Eq.~(\ref{eq7}) is more general, and can be applied to higher ionic concentrations, where steric effects must be taken into account.

Another interesting case is the limit of small association, $\phi\exp(\beta J)\ll1$, where the square root term in Eqs.~(\ref{eq7}) and (\ref{eq4}) can be Taylor-expanded, yielding $n_{s}=n_{b}\left[1-\phi\exp(\beta J)\right]\simeq n_{b}$. As expected, almost all ions are free in this case, due to a combination of low ionic strength and small association energy. In the opposite limit of large association, $\phi\exp(\beta J)\gg1$, Eq.~(\ref{eq7}) leads to $n_p a^3=\phi\left(1-\phi\right)$ and $n_{s}a^{3}=\exp\left(-\beta J/2\right)\sqrt{\phi\left(1-\phi\right)}$, {\it i.e.}, the free-ion concentration is damped by a factor of $\exp\left(-\beta J/2\right)$.

The free-ion concentration, $n_s$, with and without steric effects (Eqs.~(\ref{eq7}) and (\ref{eq4}), respectively), is presented in Fig.~\ref{fig1}. Both expressions lead to a smaller amount of free ions ($n_s<n_b$), as compared to the standard theory without pairs ($n_s{=}n_b$). We also note that pair association is more pronounced for larger $J$ values, leading to a further reduction in free ions. Steric effects become evident for non-negligible concentrations ($\phi\gtrsim 0.1$), and favor pair formation that decreases the ionic volume fraction. As $\phi$ further increases, Eq.~(\ref{eq4}) breaks down, and only the full expression of Eq.~(\ref{eq3}) remains valid. As can be seen from Fig.~\ref{fig1}a, the free-ion concentration increases with $n_{b}$ and reaches a maximal value always for $\phi=0.5$, and independent of $J$. This can be explained because of the $\phi\leftrightarrow\left(1 -\phi\right)$ symmetry of the lattice-gas model.

\subsection{Modified dipolar Poisson-Boltzmann (MDPB) equation}
\label{MDPB}

So far we discussed the properties of bulk ionic solutions where the electrostatic potential, $\psi$, and the ionic concentrations are fixed throughout the solution.
However, a charged surface  induces a spatially varying electrostatic potential $\psi(\vecr)$ and ionic concentrations. Such a potential solves Poisson's equation,
\be
\label{eq00}
-\eps_{0}\nabla^{2}\psi=\rho_{f}+\rho_{s}+\rho_{p},
\ee
where $\rho=\rho_{f}+\rho_{s}+\rho_{p}$, is the total charge density. The first term, $\rho_{f}$, is the fixed charge density of any macromolecules and surfaces, $\rho_{s}$ is that of salt ions, and $\rho_{p}$ is related to the polarization field of the solvent and Bjerrum pairs. Note that the vacuum permittivity $\eps_0$ is used, because the contribution of the solvent dipolar molecules is taken into account explicitly via $\rho_p$.
In thermal equilibrium, the densities $\rho_{s}$ and $\rho_{p}$ are further related to the electrostatic potential by their corresponding Boltzmann factor (not shown here for simplicity). Combining these terms in Poisson's equation, Eq.~(\ref{eq00}), leads to the {\it modified dipolar} PB (MDPB) equation~\cite{Borukhov00,Abrashkin07,Iglic11,Levy12,Levy13}. This equation takes into account steric effects by relating $a$, the lattice constant of Section~\ref{bulk_concentration}, with a reference close-packing density, $a^{-3}$.

Consider a planar surface with a fixed charge distribution, $\rho_{f}$, homogeneous in the $xy$-plane. The corresponding MDPB equation
depends only on the distance $z$ from the surface and reads (see Appendix):
\begin{align}
\label{eq8}
\eps_{0}\psi''(z) & =-\rho_{f}(z)+2 e n_{s}\frac{\sinh\left[\beta e\psi(z)\right]}{\mathcal{D}(z)}\nonumber\\
&- p\,n_{p}\frac{\D}{\D z}\frac{\mathcal{G}\left[\beta p\psi'(z)\right]}{\mathcal{D}(z)}\nonumber\\
&- p_{w}\left(a^{-3}-2n_{s}-n_{p}\right)\frac{\D}{\D z}\frac{\mathcal{G}\left[\beta p_{w}\psi'(z)\right]}{\mathcal{D}(z)},
\end{align}
where $\psi'=\D \psi/ \D z$ and the first two terms on the right-hand-side are $\rho_{f}$ and $\rho_{s}$, while the last two are the contributions of dipolar Bjerrum pairs and  solvent molecules to $\rho_{p}$, respectively. In Eq.~(\ref{eq8}), $\mathcal{G}(u)=\cosh u/u-\sinh u/u^{2}$ is related to the Langevin function $\mathcal{L}(u)=\coth u-1/u$ by $\mathcal{G}=\left(\sinh u / u\right)\mathcal{L}$. This is the polarization density, written as a product of the dipole density, $\sinh u/u$, and the average dipole moment, given on the MF level by the Langevin function, $\mathcal{L}$.

The denominator in Eq.~(\ref{eq8}), $\mathcal{D}$, is given by
\begin{align}
\label{eq9}
\mathcal{D}(z)&=2n_{s}a^{3}\cosh\left(\beta e\psi\right)+n_{p}a^{3}g\left(\beta p\psi'\right)\nonumber\\
&+\left(a^{-3}-2n_{s}-n_{p}\right)a^{3}g\left(\beta p_{w}\psi'\right),
\end{align}
where the function $g(u)=\sinh u/u$ satisfies $g'(u)=\mathcal{G}(u)$. The denominator, $\mathcal{D}$, leads to a saturation of the ionic and dipolar concentrations, bounded from above by the close-packing density, $a^{-3}$. This saturation feature is especially important near highly charged surfaces that attract counterions and dipoles, as is demonstrated in Section~\ref{sec4}.

Equation~(\ref{eq8}), together with the relations between $n_{s}$ $n_{p}$, and $n_{b}$, define our model for Bjerrum pairs in ionic solutions. This model can be obtained formally by employing a field-theoretical approach to a lattice-gas model including electrostatic interactions, as is further explained in the Appendix~\ref{appB}. Within this framework, Eq.~(\ref{eq8}) is the Euler-Lagrange equation, derived from the variation of the free energy. \ra{Note that in the formulation of this section as well as in the Appendix, we account only implicitly for some of the Coulombic interactions in the system via the association energy, $J$, and the hydrated ionic diameter, $a$ (see Section~\ref{bulk_concentration}).}

\ra{Equation~(\ref{eq8}) can be simplified under several conditions. For small electrostatic potentials and fields, it is possible to linearize the equation, as is further discussed in Section~\ref{sec3}. This is especially relevant in the bulk electrolyte, far from any fixed charge density. }

In addition, as opposed to the picture of Section~\ref{bulk_concentration}, where solvent molecules are considered explicitly, it is possible to treat the solvent implicitly as a homogeneous background with a dielectric constant $\eps$ of the solution, by inserting $p_{w}=0$ and substituting $\eps_{0}\to\eps$ in Eq.~(\ref{eq8}). Hereafter, we distinguish between the {\it explicit solvent} model of Eq.~(\ref{eq8}) and the {\it implicit solvent} model, described here.
Throughout this work, we present results for both models and compare between the two.

Finally, for small volume fractions of ions and Bjerrum pairs ($\phi\ll1$), it is possible to neglect steric effects. Then, the denominator function, Eq.~(\ref{eq9}), simplifies to unity, $\mathcal{D}\simeq 1$~\cite{Abrashkin07,Levy12,Levy13}, and the bulk concentrations of free ions and Bjerrum pairs are given by Eq.~(\ref{eq4})~\cite{Zwanikken09}.

\section{Dielectric constant and screening length}
\label{sec3}

The association of Bjerrum pairs modifies the permittivity of the solution and decreases the number of free ions responsible for screening. \ra{As both the dielectric constant and screening length characterize the bulk electrolyte,} the above effects can be captured by linearizing Eq.~(\ref{eq8}), leading to the general form
\begin{align}
\label{eq14b}
\eeff\psi''=2\beta e^2 n_s\psi,
\end{align}
where $\eeff$ is the effective dielectric constant, given by
\begin{align}
\label{eq16}
\eeff(\phi)
&=\eps_{0}+\frac{1}{3}\beta\left[\left(a^{-3}-2n_{s}(\phi)\right)p_{w}^{2}+n_{p}(\phi)\left(p^{2}-p_w^2\right)\right].
\end{align}
This expression is typical of effective medium theory, where the contribution of each species is proportional to its volume fraction in the solution.

For pure solvent, Eq.~(\ref{eq16}) yields $\eps=\eps_{0}+\beta a^{-3}p_{w}^{2}/3$. This is a known MF result for a dilute phase of dipoles and is not expected to yield the correct dielectric constant of pure water. For example, substituting $a=0.5\,$nm, the value $\eps=78$ corresponds to $p_{w}\approx9.8\,{\rm D}$, which is about five times larger than the physical value $p_{w}=1.85\,{\rm D}$ of water molecules. Therefore, in addition to $a$, the dipole moment $p_w$ should be regarded as a parameter of the model. A more physical relation between $p_w$ and $\eeff$ can be obtained by incorporating electrostatic correlations and fluctuations beyond MF~\cite{Levy12,Levy13}.

In the absence of pairs ($n_p=0$), Eq.~(\ref{eq16}) describes a dielectric decrement with increased ionic concentration that is known for aqueous solutions~\cite{Hasted48,Barthel95}. In Eq.~(\ref{eq16}), the decrement originates from the \ra{condition of Eq.~(\ref{eq0}).} An increased ionic concentration results in a decreased solvent concentration, reducing the effective permittivity. Moreover, ions in an aqueous solution are surrounded by an hydration shell of water molecules that are not free to rearrange themselves in response to an external field, thus lowering the dielectric constant~\cite{Dan11}.

Ion-pair association makes the above picture more complex. Bjerrum pairs have a permanent dipole moment that increases the dielectric constant. Each pair adds to the dielectric constant a term proportional to $p^2-p_{w}^2$, and the behavior of $\eeff(\phi)$, therefore, changes qualitatively according to the sign of the difference $p-p_w$. This statement is further discussed below.

The effective screening length, $\leff=1/\keff$, is obtained from Eq.~(\ref{eq14b}) by converting it to the form $\psi''=\keff^2\psi$. This leads to
\be
\label{eq15}
\ra{\leff(\phi)=\frac{1}{\keff(\phi)}=\sqrt{\frac{\eeff(\phi)}{2 \beta e^{2}n_{s}(\phi)}}.}
\ee
While it is similar to the Debye screening length of Eq.~(\ref{ld}), the above Eq.~(\ref{eq15}) suggests a more intricate dependence on the concentration $\phi$ as both  $n_{s}=n_s(\phi)$ and $\eeff=\eeff(\phi)$. In particular, the screening length does not necessarily \ra{decrease} monotonically with $\phi$~\cite{Zwanikken09}, as opposed to the classical DH result.

In order to explore this possibility, we examine the derivative of $\keff^2$ with respect to $\phi$
\be
\label{eq16a}
\frac{1}{\keff^2}\frac{\partial \keff^2}{\partial \phi}=\frac{1}{n_s} \frac{\partial n_s}{\partial \phi}-\frac{1}{\eeff}\frac{\partial \eeff}{\partial \phi}.
\ee
The first term on the right-hand-side of Eq.~(\ref{eq16a}) describes whether the free-ion concentration increases or reduces with $\phi$. According to Sec.~\ref{bulk_concentration}, it is positive for $\phi<1/2$ and negative for $\phi>1/2$. The second term, on the other hand, depends on whether ions induce a dielectric decrement or increment. As described above, this greatly depends on how large is the ion-pair dipole in comparison to the solvent dipole.

Below, we distinguish between the two cases, $p<p_w$ and $p>p_w$, and examine the behavior of $\eeff(\phi)$ and $\keff(\phi)$  for both cases. \ra{For aqueous solutions, $p<p_w$ due to the high dielectric constant of water. In order to examine the $p>p_w$ scenario,}  we broaden the scope of our discussion to general solvents rather than focusing exclusively on water.
\begin{figure*}[ht]
\centering
\begin{subfigure}[b]{0.45\textwidth}
\includegraphics[width=0.85\textwidth]{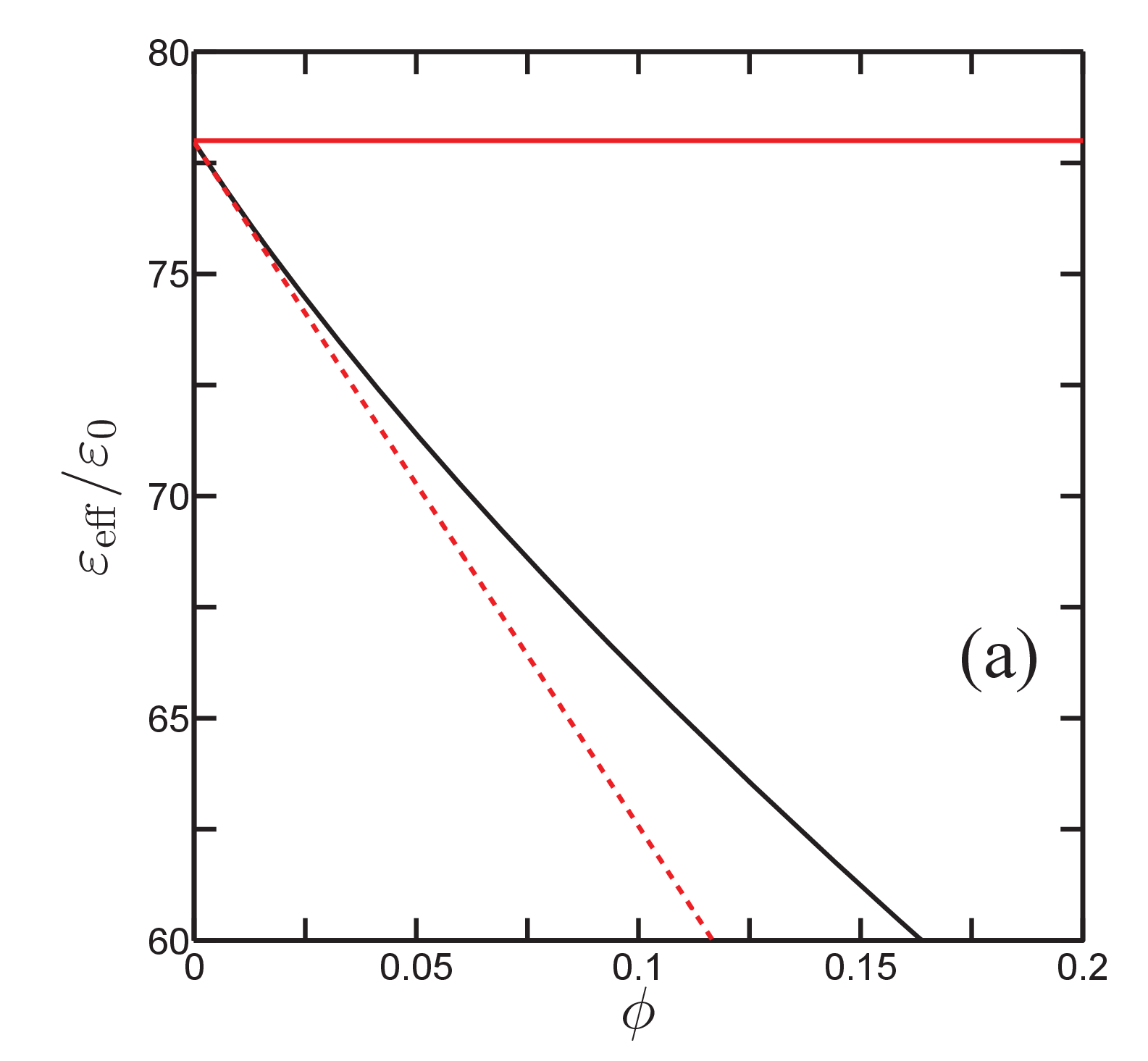}
\end{subfigure}
\begin{subfigure}[b]{0.45\textwidth}
\includegraphics[width=0.85\textwidth]{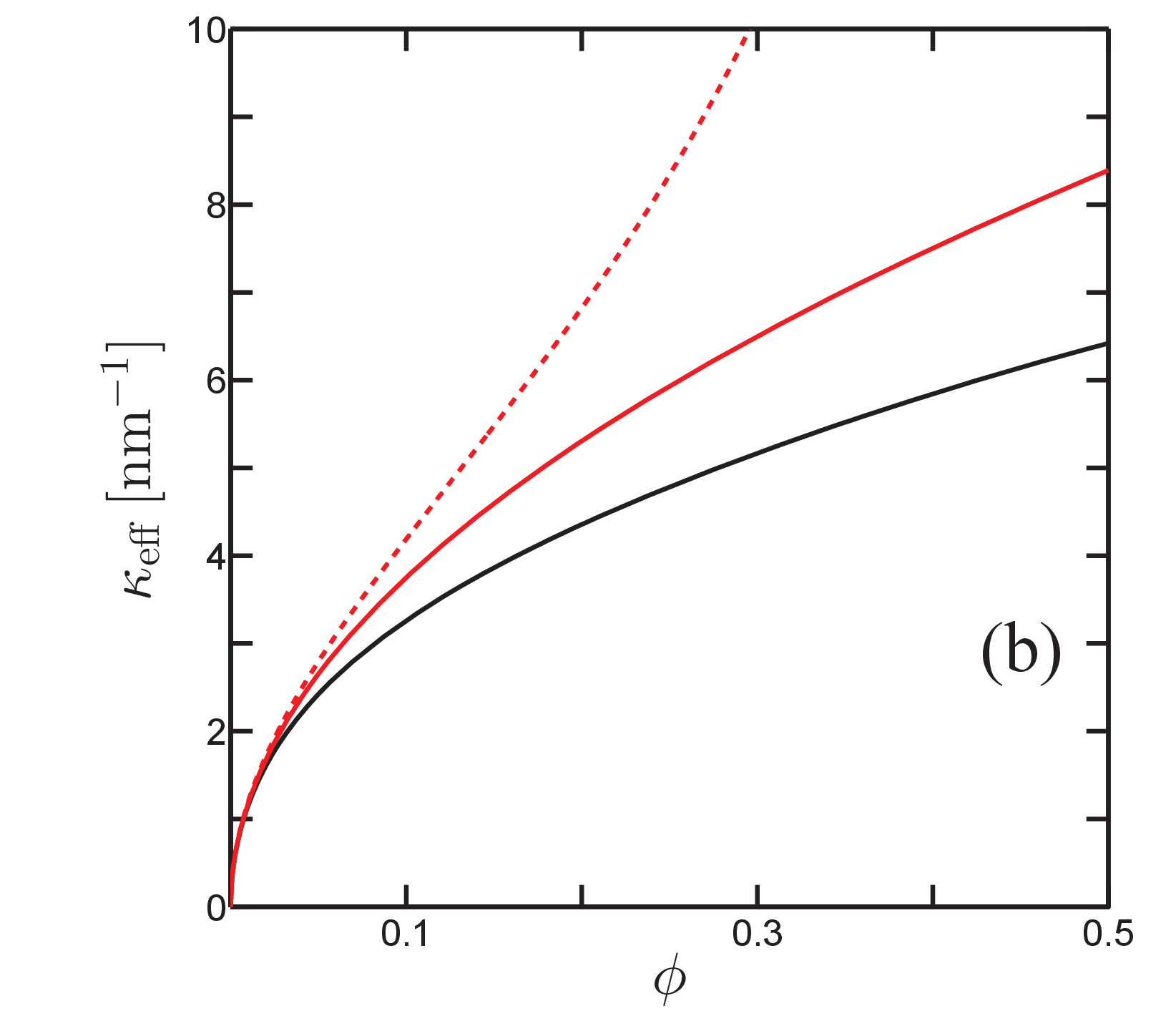}
\end{subfigure}
\caption{(a) Effective dielectric constant, $\eeff$, and (b) inverse screening length, $\keff$, as function of the dimensionless bulk ionic concentration $\phi=n_b a^3$ for $p<p_w$. The result of Eq.~(\ref{eq15}) is plotted as a solid black curve and is compared to the classical DH theory (solid red line) and to Eq.~(\ref{eq15}) without Bjerrum pairs ($n_p=0$, dashed red curve). The curves are plotted for $J=2\kbt$, $a=0.5$ nm, $b=0.1$ nm. We also use $p_{w}=9.8\,{\rm D}$ in order to fit the dielectric constant of water, $\eps=78\,\eps_0$.  }
\label{fig2}
\end{figure*}

\subsection{The $p_w>p$ case}
For $p_{w}>p$, ion pairs decrease the dielectric constant and Eq.~(\ref{eq16}) captures the dielectric decrement described after Eq.~(\ref{eq16}) for aqueous solutions.  This is illustrated in Fig.~\ref{fig2}a. At low $\phi$ values, $n_p$ is still small, and the dielectric constant decreases linearly with $n_s$, similarly to substituting no pairs ($n_p=0$ and $n_s a^3=\phi$) in Eq.~(\ref{eq16}), which is drawn as the dashed red curve in Fig.~\ref{fig2}a. For higher ionic concentrations, Bjerrum-pair dipoles contribute to the permittivity, resulting in a smaller decrease (black curve in Fig.~\ref{fig2}a). The diminished decrement for high ionic concentrations is consistent with measurements and was previously accounted for theoretically by electrostatic correlations beyond MF~\cite{Levy12,Levy13}.

The $\keff(\phi)$ behavior is illustrated in Fig.~\ref{fig2}b. As the dielectric constant decreases in our model, $\keff\sim1/\eeff$ is expected to be larger than in the classical DH theory. This is indeed the case in the absence of pairs (red curves in Fig.~\ref{fig2}b). However, the shallow decrement in presence of Bjerrum pairs, alongside the smaller number of free ions, result in a $\keff$ (black curve) that is smaller than the DH result. Furthermore, $\keff$ increases monotonically, in accordance with Eq.~(\ref{eq16a}). Note that $\keff(\phi)$ can decrease for $\phi$ values larger than $1/2$ (not shown in Fig.~\ref{fig2}b), where fluctuations are expected to play an important role and the MF approximation breaks down.
%

\begin{figure*}[ht]
\centering
\begin{subfigure}[b]{0.45\textwidth}
\includegraphics[width=0.85\textwidth]{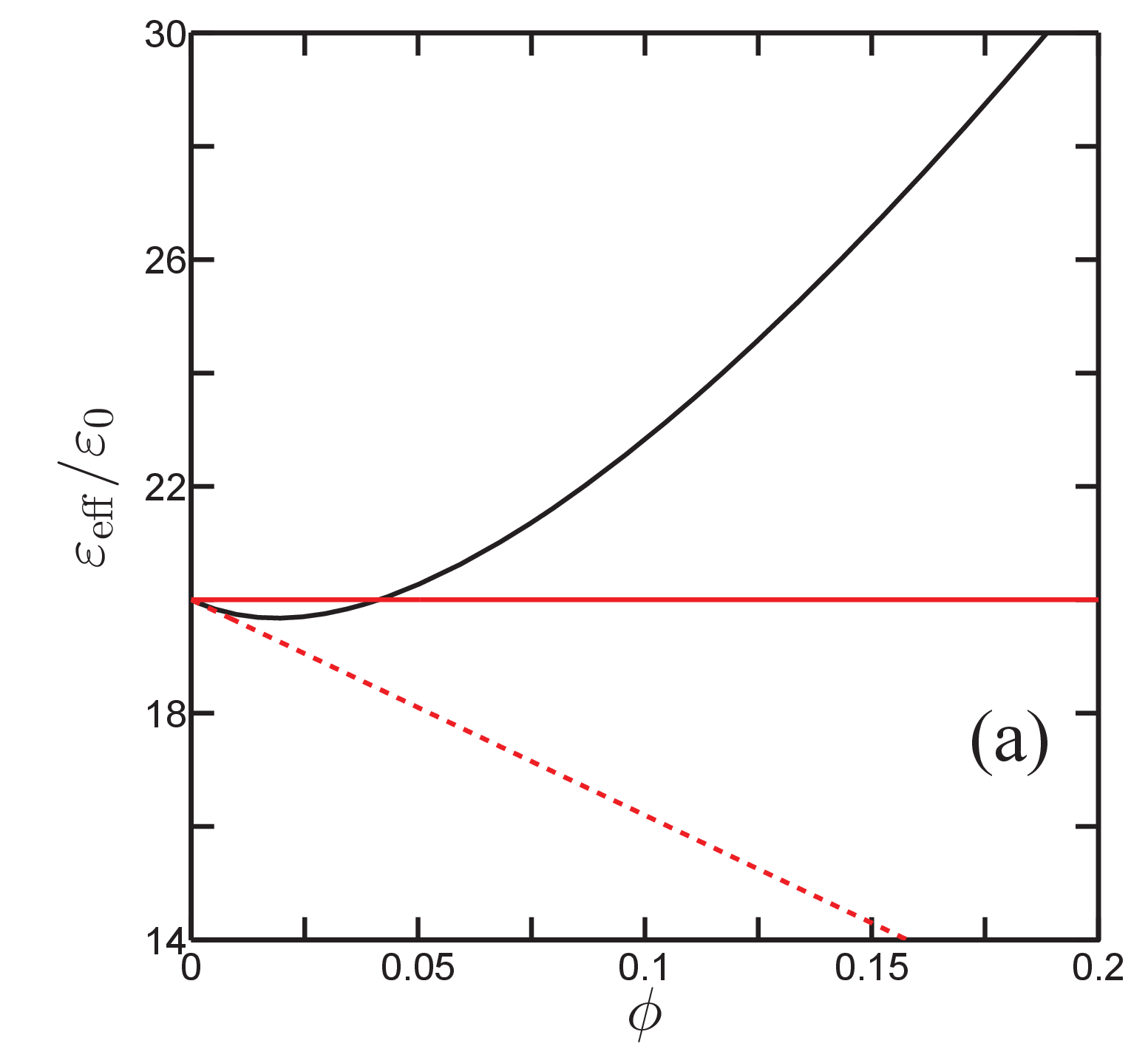}
\end{subfigure}
\begin{subfigure}[b]{0.45\textwidth}
\includegraphics[width=0.85\textwidth]{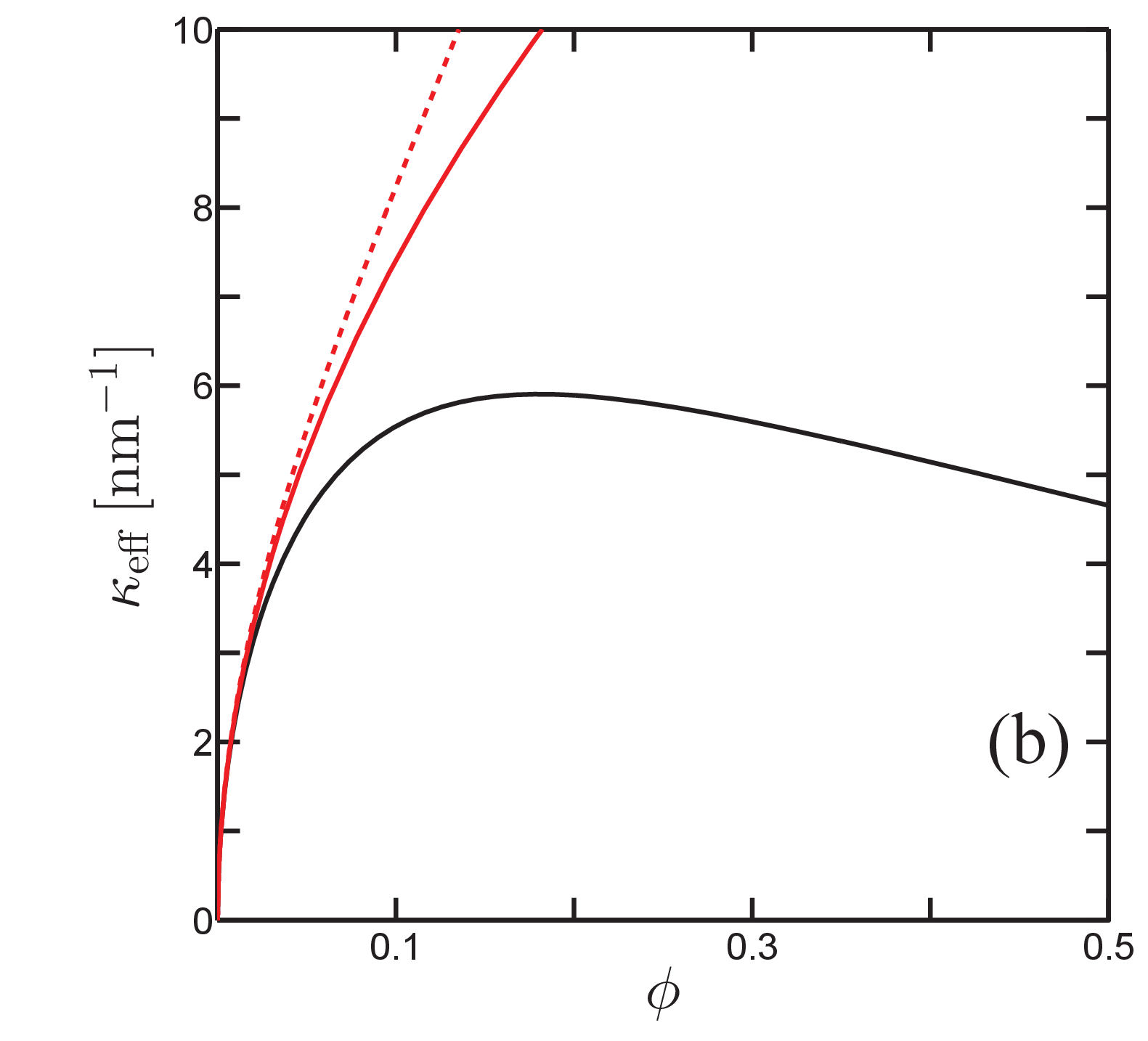}
\end{subfigure}
\caption{(a) Effective dielectric constant, $\eeff$, and (b) inverse screening length, $\keff$, as function of the dimensionless bulk ionic concentration $\phi=n_b a^3$ for $p>p_w$. The result of Eq.~(\ref{eq15}) is plotted as a solid black curve and is compared to the classical DH theory (solid red line) and to Eq.~(\ref{eq15}) without Bjerrum pairs ($n_p=0$, dashed red curve). The curves are plotted for $J=2\kbt$, $a=0.5$ nm, $b=0.3$ nm. The dipolar moment $p_{w}=4.9\,{\rm D}$ is used in order to fit the dielectric constant, $\eps=20\,\eps_0$. \ra{We emphasize that these figures are not appropriate to aqueous solutions, where $p<p_w$.}  }
\label{fig3}
\end{figure*}

\subsection{The $p_w<p$ case}
We consider a solvent such as alcohol with $\eps=20$, smaller than that of water. For such relatively low $\eps$ values, consistent with $p_w<p$, we find that the dielectric constant is non-monotonic with $\phi$, as is illustrated in Fig.~\ref{fig3}a. While at low $\phi$ values, $\eeff$ follows the linear decrement of the theory without pairs (dashed red curve), the Bjerrum-pair contribution at higher concentrations leads to a dielectric {\it increment} (black curve). Such an increment can also be seen in the implicit solvent model, where $\varepsilon_{\rm eff} = \varepsilon + \beta n_p p^2/3$, and $\varepsilon$ is the dielectric constant of the solution without the ionic pairs.

The $\keff(\phi)$ behavior in the $p_w<p$ case is smaller than the result of classical DH theory as well as of Eq.~(\ref{eq15}) without pairs. However, the screening length in Fig.~\ref{fig3}b is non-monotonic. We emphasize that this anomalous behavior of $\keff(\eps)$ originates from the dielectric increment [Eq.~(\ref{eq16a})], and is {\it inconsistent} with the dielectric decrement that was measured for aqueous solutions.

\section{Ionic profiles close to a charged plate}
\label{sec4}

Beside exploring bulk properties of ionic solutions, we also investigate how ion-pair formation influences the ionic concentration near a charged surface. Consider the ionic solution of Section~\ref{sec2} confined to the $z{>}0$ half-space by a charged surface at $z{=}0$. The surface is homogeneously charged with a surface-charge density, $\sigma$, {\it i.e.,} $\rho_{f}(z)=\sigma\delta(z),$ where $\delta(z)$ is the Dirac delta function.

The boundary condition at $z{=}0$ is determined by Gauss' law by integrating Eq.~(\ref{eq8}) over a small interval around $z{=}0$. We assume that the electric field is confined to the upper region $z{>}0$, leading to
\begin{align}
\label{eq10}
-\eps_{0}\psi'(0)&=\sigma+pn_{p}\frac{\mathcal{G}\left[\beta p\psi'(0)\right]}{\mathcal{D}(0)}\nonumber\\
&+p_{w}\left(a^{-3}-2n_{s}-n_{p}\right)\frac{\mathcal{G}\left[\beta p_w\psi'(0)\right]}{\mathcal{D}(0)}.
\end{align}
The boundary condition is more complex than in the regular PB model and includes the free surface-charge density, $\sigma$, as well as the polarization-induced bound surface-charge density, which lowers (in absolute value) the total surface charge .

\begin{figure}[ht]
\centering
\includegraphics[width=0.95\columnwidth]{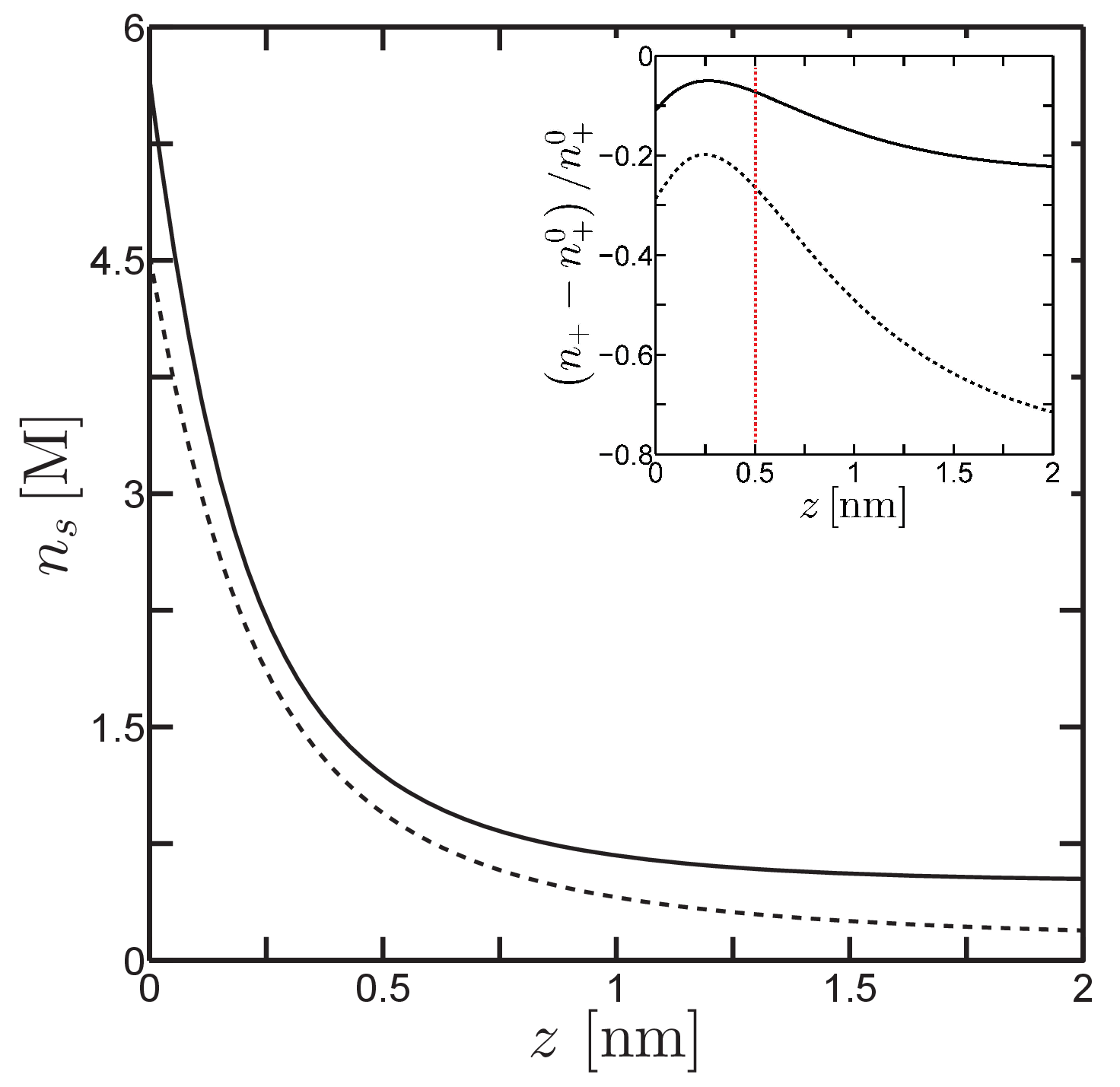}
\caption{Counterion concentration profiles, $n_{+}(z)$, for $J=2\kbt$ (solid) and $J=6\kbt$ (dashed). All the profiles are plotted for the implicit solvent model with $\lb=0.7$ nm, $a=b=0.5$ nm, $|\sigma|=e/\,{\rm nm}^{2}$, and $\phi=0.05$. These values correspond to $n_b=0.66$\,M, $n_s=0.51$\,M for $J=2\kbt$, and $n_s=0.13$\,M for $J=6\kbt$. The inset shows the relative difference as compared to the $n_{+}^{0}$ profiles of the theory without pairs. The lattice size, $a$, is marked in the inset by a dotted red vertical line.
}
\label{fig4}
\end{figure}

Once Eq.~(\ref{eq8}) is solved and the electrostatic potential is found, the ionic profiles $n_{\pm}(z)$ as function of the distance from the surface, $z$, are evaluated from the corresponding Boltzmann factor,
\be
\label{eq13a}
n_{\pm}(z)=n_{s}\frac{\e^{\mp \beta e \psi(z)}}{\mathcal{D}(z)}.
\ee
As the boundary condition of Eq.~(\ref{eq10}) is too complex to be solved analytically, we present in Fig.~\ref{fig4} numerical results for the counterion profile, $n_{+}(z)$.
For simplicity, we treat the solvent {\it implicitly} as a homogeneous dielectric background with a dielectric constant $\eps$, as opposed to dipoles in vacuum. This is equivalent to substituting $p_w\to0$ and $\eps_0\to\eps$  in Eqs.~(\ref{eq16}) and (\ref{eq10}), as well as in Eqs.~(\ref{eq8}) and (\ref{eq9}) in Section~\ref{MDPB}, and
is applicable at low $\phi$ values.

\ra{Counterion profiles of positive ions next to a negatively charged surface are illustrated in Fig.~\ref{fig4} for two $J$ values. The effect of ion association is evident from the inset, where the relative difference, as compared to the theory without ion pairs, is presented. We find that the counter\-ion concentration is overall smaller due to the lower bulk value, $n_s$, that decreases as function of $J$ (Section~\ref{bulk_concentration}).}

\ra{Furthermore, both free ions and ion pairs accumulate next to the charged surface, where the electrostatic potential and field are the strongest. Although free ions are attracted more strongly to the surface, pairs are also present at the surface proximity due to mixing entropy.  As a result of this accumulation, steric effects are pronounced up to a distance $a$ from the surface, as is illustrated in Fig.~\ref{fig4}.} Far away from the surface, the counterion concentration reaches its bulk value, $n_s$.

\section{Osmotic pressure between two charged plates}
\label{sec5}

We consider two charged surfaces at $z=\pm d/2$, bounding an ionic solution. The top surface has a surface-charge density $\sigma_1$, and the lower $\sigma_2$. We focus on the scenario of oppositely charged surfaces, $\sigma_1=-\sigma_2=\sigma$, where the ion-pair formation is expected to have the most evident effect on the osmotic pressure. The electric field is larger between oppositely charged surfaces, as compared to equally charged ones, leading to higher dipolar concentration and larger deviation from the standard PB theory. As in Section~IV, we assume that the electric field is confined to the aqueous solution region, $-d/2<z<d/2$.

The osmotic pressure, $\Pi=p_{{\rm in}}-p_{{\rm out}}$, is defined as the difference between the pressure inside the solution and the one exerted by the bulk electrolyte reservoir. The inner pressure, constant throughout the system, is equal to the first integral~\cite{Dan09,Safinyabook} of the differential equation, Eq.~(\ref{eq8}). The outer pressure is obtained by setting the electrostatic potential and electric field to zero. Taking the first integral of the MDPB equation, Eq.~(\ref{eq8}), and subtracting the outer pressure leads to an exact expression,
\begin{align}
\label{eq13}
\Pi&=-\half\psi'\left[\eps_{0}\psi'+2pn_{p}\frac{\mathcal{G}(\beta p\psi')}{\mathcal{D}}\right.\nonumber\\
&+\left.2p_{w}\left(a^{-3}-2n_{s}-n_{p}\right)\frac{\mathcal{G}(\beta p_{w}\psi')}{\mathcal{D}}\right]+\kbt a^{-3}\ln\mathcal{D}.
\end{align}
In the above equation, the attractive (negative) term originates from the electrostatic energy density $\sim\mathbf{E}\cdot\mathbf{D}$, with $\mathbf{E}=-\psi'\hat{z}$ being the electric field, and $\mathbf{D}=D\hat{z}$ the displacement field. The repulsive (positive) terms, on the other hand, originate from the mixing entropy of ions and of dipoles as well as the rotational entropy of dipoles $\sim \ln\mathcal{D}$.

\begin{figure}[ht]
\centering
\includegraphics[width=0.95\columnwidth]{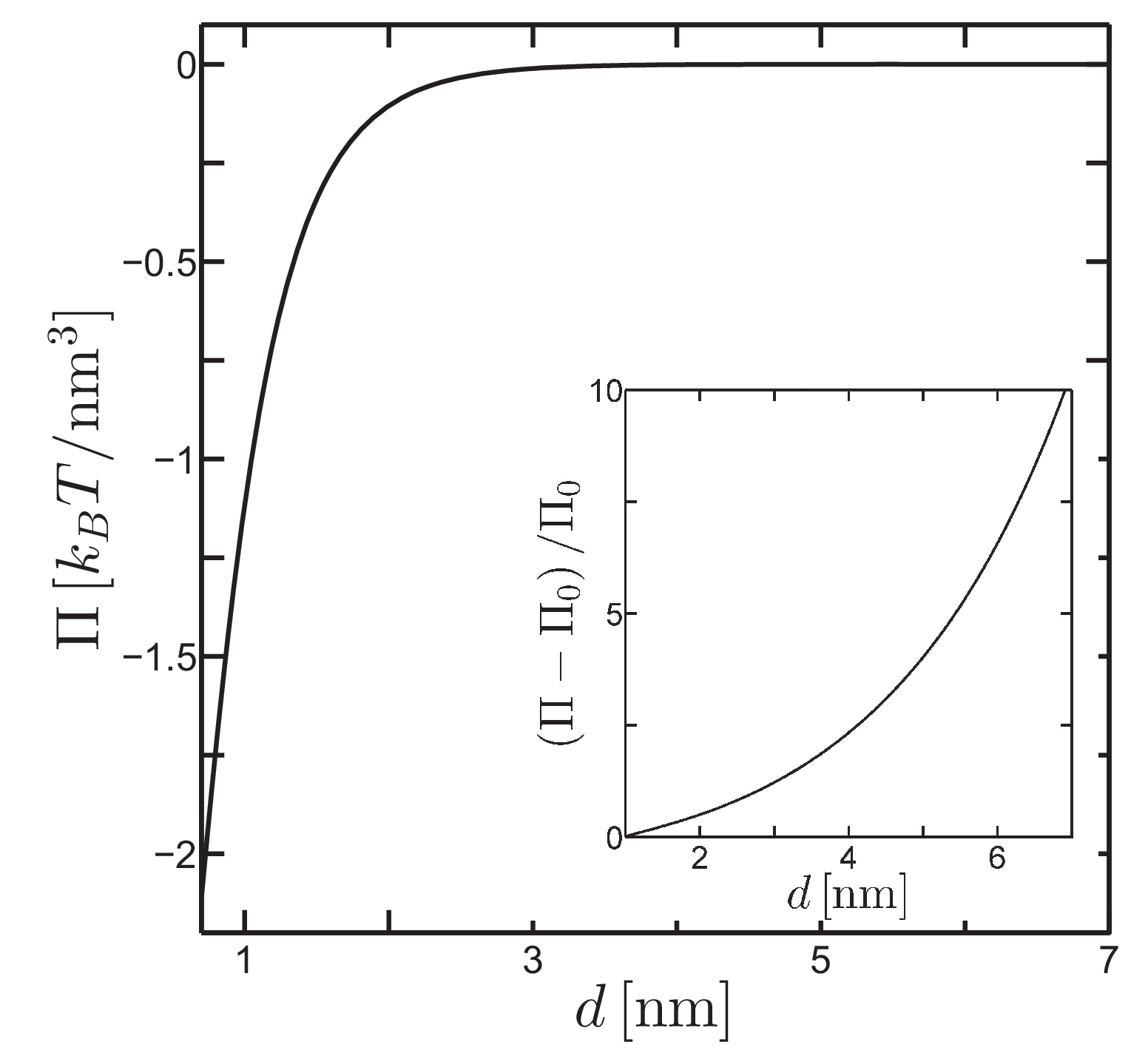}
\caption{Osmotic pressure profile between oppositely charged surfaces for $J=2\kbt$, $\lb=0.7$ nm, $a=b=0.5$ nm, $|\sigma|=e/\,{\rm nm}^{2}$, and $\phi=0.05$. The inset shows the relative difference with respect to the $\Pi_{0}$ pressure (no pairs). For simplicity, the solvent is treated implicitly.}
\label{fig5}
\end{figure}

The electrostatic potential can be solved numerically from Eq.~(\ref{eq8}), with the boundary conditions that are obtained in Section~\ref{sec4} [Eq.~(\ref{eq10})], and applied to the two surfaces at $z=\pm d/2$. The osmotic pressure is then found via Eq.~(\ref{eq13}). A characteristic pressure profile between oppositely charged surfaces is illustrated in Fig.~\ref{fig5}.

\ra{The pressure in Fig.~\ref{fig5} is negative due to the electrostatic attraction between oppositely charged surfaces. It is enhanced  by the Bjerrum pairs, as is evident from the relative difference with respect to the theory without pairs, illustrated in the inset. This difference becomes large by an order of magnitude at large separations, where the electrostatic potential and field become small and the linear approximation of Section~\ref{sec3} is valid. The linearized form leads to an exponentially decaying osmotic pressure, $\Pi\sim-\exp\left(-\keff d\right)$, with a  screening length that decreases due to ion association (Section~\ref{sec3}). The relative difference, therefore, is expected to increase exponentially.}

We note that for sufficiently small inter-surface separations, the two surfaces effectively neutralize each other, and free ions are expected to be released to the bulk~\cite{Sam05,Dan07} for entropic gain. In such a case, screening effects are negligible, and the above reduction in the pressure should not necessarily hold.

\section{Discussion}
\label{sec6}

The present work addresses the effect of Bjerrum pairs on the properties of ionic solutions in the bulk and near charged surfaces. The main result of our model is the MDPB equation in presence of the Bjerrum pairs, Eq.~(\ref{eq8}). This MF equation is written in terms of the free-ion and pair densities, $n_{s}$ and $n_{p}$, respectively, determined by the lattice-gas model of Section~\ref{bulk_concentration}. As pairs are predicted to associate for relatively high ionic concentrations, the theory accounts for steric effects and the dielectric decrement of ionic solutions, which are important under such physical conditions.

The ionic mixture considered in our model is symmetric, composed of anions and cations of the same valency and comparable size. However, asymmetric mixtures can also be studied in a similar fashion. For ions and dipoles of different sizes, it is possible to employ an asymmetric lattice-gas model (see, e.g., Ref.~\cite{maggs16}). Furthermore, for anions and cations of different valencies, one can consider the association into a charged anion-cation pair~\cite{dosSantos10}.

The model includes four physical parameters: the association energy $J$, lattice size $a$, the Bjerrum dipolar moment $p$, and the water dipolar moment $p_{w}$.
These parameters can be evaluated from measurements of several physical quantities, such as the dielectric constant. While we treat these parameters as independent, they are in fact coupled. Namely, the values of the  $J$, $p$, and $p_{w}$ parameters depend on the lattice size, $a$. Furthermore, the quantities are coupled via the effective solvent permittivity that decreases the association energy and increases the hydration shell. Such mutual dependence can be  taken into account to some extent via the Bjerrum postulate~\cite{Bjerrum26}, according to which the kinetic constant $K$ of pair association, $K=\left[n_{s}\right]^{2}/\left[n_{p}\right]$, is proportional to the inverse of the integral $\int_{v}\D\vecr\exp\left(\lb/r\right)$. The integration volume, $v$, is the region within which the ions are considered associated, related to the ionic diameter and Bjerrum length~\cite{Bjerrum26,Fisher93,Levin96,Zwanikken09}.

Recent surface-force experiments~\cite{Smith16} suggest a non-monotonic dependence of the screening length on the bulk ionic concentration for ionic concentrations corresponding to a range of $\phi$ values as low as $0.1$. This behavior was recently attributed to solvent molecules acting as defects in an ionic crystal~\cite{Lee17}. Our prediction for the effective screening length [Eq.~(\ref{eq15})] deviates from the usual PB result. We obtain different results, depending on the relative dipolar moment $p-p_w$. While for $p<p_w$, the screening length is monotonically decreasing, it can increase with ionic concentration for $p>p_w$. For aqueous solutions and simple salt, the $p>p_w$ scenario is \ra{not appropriate}, and expected for lower dielectrics (e.g., alcohol) and/or polarizable ions. We demonstrate in Eq.~(\ref{eq16a}) that in light of the dielectric decrement of ionic solutions, the screening length can increase with the ionic concentration only due to a reduction in the amount of free ions.

The reduction of free-ion concentration at low $\phi$ values can have two origins. Steric effects can be more important than what the lattice-gas framework conveys. For example, it is  possible to take steric effects into account by using a virial expansion of hard spheres, where the maximal value of the free ionic concentration, $n_{s}$, is calculated at $\phi\approx 0.1$~\cite{Sam17}. In addition, correlations beyond MF between ions and dipoles have a crucial role in highly concentrated ionic solutions and ionic liquids, and may promote the reduction of $n_{s}$. This possibility will be explored in the future by employing a loop expansion of the system free-energy, leading to predictions beyond the MF theory.

\vskip 0.5cm
{\it Acknowledgments.~}
This work was motivated by discussions with S. Safran. We thank him and R. Podgornik for numerous helpful suggestions and comments. This work was supported by the Israel Science Foundation (ISF) under grant No. 438/12, the U.S.- Israel Binational Science Foundation (BSF) under grant No. 2012/060, and the ISF-NSFC joint research program under grant No. 885/15. D.A. thanks Alexander von Humboldt Foundation for a Humboldt research award, and T.M. acknowledges support from the Blavatnik postdoctoral fellowship programme.

\appendix*
\section{Field-theoretical approach}
\label{appB}

Consider the aqueous solution of Section~\ref{sec2} modeled via a cubic lattice with
cells of size $a$.
Each cell is occupied either by a solvent molecule with dipole moment $p_{w}$, a cation of charge $e$, an anion of charge $-e$, or a Bjerrum pair with dipole moment $p$.
The association energy of pairs is $-J,$ where $J$ is a positive energy parameter.

The partition function can be written in terms of spin-like
variables assigned to each cell, accounting for its occupation. Each
cell $j,$ located at $\boldsymbol{r}_{j},$ is described by a pair of variables, $\left(s_{j}^{+},s_{j}^{-}\right),$ counting the cations and anions
in the cell, respectively, where $s_{j}^{\pm}=0\,{\rm or}\,1$. With these variables, the charge density operator is given by
\begin{align}
\label{eqa1}
\hat{\rho}(\vecr) & =\rho_{f}(\vecr)+e\sum_{j}\left[\left(s_{j}^{+}-s_{j}^{-}\right)\delta\left(\vecr-\vecr_{j}\right)\right.\nonumber\\
&-\left.\left[s_{j}^{+}s_{j}^{-}p+\left(1-s_{j}^{+}\right)\left(1-s_{j}^{-}\right)p_{w}\right]
\boldsymbol{\hat{n}}_{j}\cdot\boldsymbol{\nabla}\delta\left(\vecr-\vecr_{j}\right)\right],
\end{align}
where $\rho_{f}(\vecr)$ (the first term) describes any possible fixed charge
density, the 2nd term corresponds to the cations and anions, and the 3rd term accounts either for the presence of a water dipole or
a Bjerrum-pair dipole in the $j$-cell, pointing in the direction of the unit vector, $\boldsymbol{\hat{n}}_{j}$.

Incorporating the Coulombic energy and the ion-association energy yields the following
grand-partition function:
\begin{align}
\label{eqa2}
Z & =\sum_{s_{j}}\prod_{j}\exp\left[\beta\left(\mu_{+}s_{j}^{+}+\mu_{-}s_{j}^{-} + s_{j}^{+}s_{j}^{-}\right)\right]\nonumber \\
&\times\int\frac{\D\Omega_{j}}{4\pi}\exp\left[-\frac{\beta}{2}\int \D\boldsymbol{r}\D\boldsymbol{r'}\hat{\rho}\left(\boldsymbol{r}\right)v_{c}\left(|\boldsymbol{r}-\boldsymbol{r'}|\right)\hat{\rho}\left(\boldsymbol{r}'\right)\right],
\end{align}
where $\mu_{\pm}$ is the chemical potential of the positive and negative ions, respectively, $\beta=\left(k_{B}T\right)^{-1}$ is the inverse thermal energy,
$\Omega_{j}$ is the solid angle of $\boldsymbol{\hat{n}}_{j}$, and $v_{c}\left({r}\right)=1/\left(4\pi\eps_{0}\left|{r}\right|\right)$
is the Coulomb interaction. For sake of convenience, we have set the chemical potential of solvent molecules to be zero. The Coulombic self-energy can be formally
absorbed in the chemical potentials and $J$, overcoming the problematic divergence
of $v_{c}\left(0\right).$

We replace the occupation degrees of freedom, $\{s_j\}$, with a spatially fluctuating field by introducing a density field $\rho\left(\boldsymbol{r}\right)$
and its conjugate field, $\varphi\left(\boldsymbol{r}\right),$ via
the functional identity:
\begin{align}
\label{eqa3}
1 & =\int\mathcal{D}\rho\,\delta\left[\rho\left(\boldsymbol{r}\right)-\hat{\rho}\left(\boldsymbol{r}\right)\right]\nonumber \\
 & =\int\mathcal{D}\rho\mathcal{D}\varphi\exp\left(i\beta\int \D\boldsymbol{r}\varphi\left(\boldsymbol{r}\right)\left[\rho\left(\boldsymbol{r}\right)-\hat{\rho}\left(\boldsymbol{r}\right)\right]\right).
\end{align}
Substituting Eq.~(\ref{eqa3}) in Eq.~(\ref{eqa2}) and after further manipulations (for more details, see Ref.~\cite{Borukhov00}), the partition function can be written as
\begin{align}
\label{eqa6}
Z & =\int\mathcal{D}\varphi\,\e^{-\beta S\left[\varphi\right]},
\end{align}
with the field action
\begin{align}
\label{eqa7}
S\left[\varphi\right] & =\int \D\vecr\left[\frac{\eps_{0}}{2}\left(\nabla\varphi(\vecr)\right)^{2}+i\rho_{f}(\vecr)\varphi(\vecr)\right.\nonumber \\
 & -\frac{k_{B}T}{a^{3}}\ln\left({\rm sinc}\left(\beta p_{w}\left|\boldsymbol{\nabla}\varphi(\vecr)\right|\right)\right.\nonumber\\
 &+\Lambda_{+}\Lambda_{-}\e^{\beta J}{\rm sinc}\left(\beta p\left|\boldsymbol{\nabla}\varphi(\vecr)\right|\right)\nonumber\\
 &+\left.\left.\Lambda_{+}\e^{-i\beta e\varphi(\vecr)}+\Lambda_{-}\e^{i\beta e\varphi(\vecr)}\right)\right].
\end{align}

In the mean field (MF) approximation, the partition function is approximated
by its value at the saddle point, $\varphi=\varphi_{0}.$ We denote $i\varphi_{0}=\psi$,
and by using the relation $F=-k_{B}T\ln Z$, we obtain the free energy
\be
\label{eqa8}
F[\psi]=S[-i\psi].
\ee
By examining Eqs.~(\ref{eqa7})-(\ref{eqa8}) with $\Lambda_{\pm}=0$ and $p_{w}=0$, it can be seen that $\psi$ is the electrostatic potential. Furthermore, the fugacities are constant throughout the
system and are related to the total number densities of positive and negative ions, as is described in Section~\ref{bulk_concentration}.

The potential $\psi$ solves the Euler-Lagrange equation obtained from the variation $\delta F/\delta\psi=0.$ Assuming that $\rho_{f}$ depends only on the $z$-coordinate, the Euler-Lagrange equation for the free energy, Eq.~(\ref{eqa8}), coincides with Eq.~(\ref{eq8}), and is a more formal way to obtain the MDPB equation.
\newpage


\begin{thebibliography}{99}

\bibitem{Israelachvily}
J. N. Israelachvili, {\it Intermolecular and Surface Forces}, 3rd ed. (Academic, New York, 2011).

\bibitem{VO}
E. J. Verwey and J. Th. G. Overbeek, {\it Theory of the Stability of Lyophobic Colloids} (Elsevier, New York, 1948).

\bibitem{David95} D. Andelman, in {\it Handbook of Physics of Biological Systems}, edited by R. Lipowsky and E. Sackmann, Vol. I (Elsevier Science, Amsterdam, 1995), Chap. 12.
\bibitem{DH} P. Debye and E. H\"uckel, {\it Physikalische Zeitschrift} {\bf 24}, 185 (1923).
\bibitem{Bjerrum26} 13] N. Bjerrum, {\it Kgl. Dan. Vidensk. Selsk. Mat. fys. Medd.} {\bf 7}, 1 (1926).
\bibitem{Fisher93} M. E. Fisher and Y. Levin, {\it Phys. Rev. Lett.} {\bf 71}, 3826 (1993).
\bibitem{Levin96} Y. Levin and M. E. Fisher, {\it Physica A} {\bf 225}, 164 (1996).
\bibitem{Netz00} R. R. Netz and H. Orland, {\it Eur. Phys. J. E} {\bf 1}, 67 (2000).
\bibitem{Zwanikken09} J. Zwanikken  and R. van Roij, {\it J. Phys.: Condens. Matter} {\bf 21}, 424102 (2009).
\bibitem{Hasted48} J. B. Hasted, D. M. Ritson, and C. H. Collie, {\it J. Chem. Phys.} {\bf 16}, 1 (1948).
\bibitem{Barthel95} J. Barthel, R. Buchner, and M. M{\"u}nsterer, {\it Electrolyte Data Collection: Dielectric Properties of Water and Aqueous Electrolyte Solutions (Chemistry Data)}
(Dechema, Frankfurt am Main, Germany, 1995).
   \bibitem{Levy12} A. Levy, D. Andelman, and H. Orland,
{\it Phys. Rev. Lett.} {\bf 108}, 227801 (2012).
\bibitem{Levy13}  A. Levy, D. Andelman, and H. Orland,
{\it J. Chem. Phys.} {\bf 139}, 164909 (2013).

\bibitem{Smith16} A. M. Smith, A. A. Lee, and S. Perkin, {\it J. Phys. Chem. Lett.} {\bf 7}, 2157 (2016).

\bibitem{Borukhov00} I. Borukhov, D. Andelman, and H. Orland, {\it Electrochim. Acta} {\bf 46}, 221 (2000).
\bibitem{Abrashkin07} A. Abrashkin, D. Andelman, and H. Orland, {\it Phys. Rev. Lett.} {\bf 99}, 077801 (2007).

\bibitem{Iglic11} E. Gongadze, U. van Rienen, V. Kralj-Igli{\v c}, and A. Igli{\v c}, {\it Gen. Physiol. Biophys.} {\bf 30}, 130 (2011).
\bibitem{Dan11} D. Ben-Yaakov, D. Andelman, and R. Podgornik, {\it J. Chem. Phys.} 134, 074705 (2011).
\bibitem{Dan09} D. Ben-Yaakov, D. Andelman, D. Harries, and R. Podgornik, {\it J. Phys. Chem. B} {\bf 113}, 6001 (2009).
\bibitem{Safinyabook} T. Markovich, D. Andelman, and R. Podgornik, in {\it Handbook of Lipid Membranes}, edited by C. Safinya and J. R{\"a}dler (Taylor and Francis, to be published).
\bibitem{Sam05} S. A. Safran,  Europhys. Lett. 69 (2005) 826.
\bibitem{Dan07} D. Ben-Yaakov, Y. Burak, D. Andelman, and S. A. Safran, {\it Europhys. Lett.} {\bf 79}, 48002 (2007).
\bibitem{maggs16} A. C. Maggs and R. Podgornik, {\it Soft Matter} {\bf 12} 1219 (2016).
\bibitem{dosSantos10} A. P. dos Santos, A. Diehl, and Y. Levin, {\it J. Chem. Phys.} {\bf 132} 104105 (2010).
     \bibitem{Lee17} A. A. Lee, C. Perez-Martinez, A. M. Smith, and S. Perkin, unpublished (2017).
    \bibitem{Sam17} S. A. Safran, unpublished (2017).










\end{thebibliography}
\end{document}